\documentclass[runningheads]{llncs}
\usepackage[T1]{fontenc}
\usepackage{microtype}
\usepackage{graphicx}
\usepackage{comment}
\usepackage{hyperref}
\usepackage{url}
\usepackage[table]{xcolor}
\usepackage{array}  
\usepackage{listings}
\usepackage{xcolor}
\usepackage{amssymb}
\usepackage{needspace}
\usepackage{makecell}
\newcolumntype{C}{>{\centering\arraybackslash}p{1.05cm}}
\usepackage{booktabs}
\usepackage{tabularx}
\newcolumntype{Y}{>{\raggedright\arraybackslash}X}
\newcolumntype{C}{>{\centering\arraybackslash}p{1.0cm}}
\renewcommand{\arraystretch}{1.15}

\usepackage{amssymb}

\usepackage{enumitem}
\setlist{nosep,leftmargin=*,itemsep=2pt,topsep=2pt,parsep=0pt}
\setlist[itemize]{label=\textbullet}
\setlist[enumerate]{label=\arabic*.)}

\usepackage[T1]{fontenc}
\usepackage{listings}
\usepackage{xcolor}
\definecolor{codebg}{RGB}{248,248,248}
\definecolor{codeframe}{RGB}{220,220,220}
\definecolor{codenumbers}{RGB}{120,120,120}

\newcommand{\wraparrow}{\(\hookrightarrow\)}

\lstdefinestyle{base-style}{
  backgroundcolor=\color{codebg},
  frame=single,
  rulecolor=\color{codeframe},
  frameround=tttt,
  basicstyle=\ttfamily\small,
  numbers=left,
  numberstyle=\color{codenumbers}\footnotesize,
  numbersep=8pt,
  xleftmargin=1.2em,
  framesep=6pt,
  showstringspaces=false,
  breaklines=true,
  breakatwhitespace=true,
  postbreak=\mbox{\textcolor{codenumbers}{\wraparrow}\space},
  tabsize=2,
  captionpos=b,
columns=fullflexible, 
keepspaces=true,
breakindent=0pt,
linewidth=\linewidth,
basicstyle=\ttfamily\footnotesize 
}

\lstdefinelanguage{Gherkin}{
  morekeywords={Feature,Scenario,Background,Given,When,Then,And,But,Examples},
  sensitive=true,
}

\lstdefinestyle{gherkin-style}{style=base-style, language=Gherkin}
\lstdefinestyle{python-style}{style=base-style, language=Python}
\lstdefinestyle{json-style}{style=base-style, language=}
\begin{document}
\title{A Fully Automated, Deployment-Aware Testing Pipeline for IoT-Based Automotive Applications}
\titlerunning{Deployment-Aware Testing Pipeline for IoT Automotive Applications}
\author{Denesa Zyberaj\inst{1,2}\orcidID{0009-0009-4207-8723} \and\\
Roman Vintonyak\inst{1,2}\orcidID{0009-0008-1361-2205}\and
Pascal Hirmer\inst{1}\orcidID{0000-0002-2656-0095} \and
Marco Aiello\inst{2}\orcidID{0000-0002-0764-2124}}
\authorrunning{Zyberaj et al.}
\institute{Mercedes-Benz AG, Bela-Barenyi-Straße, 71059 Sindelfingen, Germany\\
\email{\{denesa.zyberaj, roman.vintonyak, pascal.hirmer\}@mercedes-benz.com}
\and
Institute for Architecture of Application Systems, University of Stuttgart,\\
Universitätsstraße 38, 70569 Stuttgart, Germany\\
\email{marco.aiello@iaas.uni-stuttgart.de}
}

\maketitle              % typeset the header of the contribution
\begin{abstract}
Testing embedded software in modern vehicles is challenging due to system complexity, decentralized architectures, and strict safety and performance constraints. In this work, we present an end-to-end, deployment-aware testing pipeline for IoT-based automotive applications. The pipeline combines requirement-driven test and code generation with large language model (LLM) and vision-language model (VLM) assistance, and human-in-the-loop curation to reduce manual effort and improve consistency. Using Eclipse openDuT, it supports flexible, distributed deployment across geographically separated cyber-physical and IoT infrastructures, optimizing for node availability and cross-organizational coordination. For validation, we conduct a case study using a Child Presence Detection System (CPDS), achieving full functional requirement coverage across all 9 requirements and 100\% Gherkin generation accuracy on the controlled requirement set. Distributed test execution across geographically separated ECUs via Eclipse openDuT confirms the pipeline's applicability to OEM--supplier testing workflows.

\end{abstract}
\section{Introduction}

The automotive industry is rapidly evolving toward software-defined, sensor-rich vehicles embedded within complex Internet of Things (IoT) ecosystems. The vehicles rely on distributed Electronic Control Units (ECUs) connected through diverse network topologies and are responsible for coordinating safety-critical functionalities~\cite{Zyberaj2024,Lami2016}. These ECUs are provisioned globally and require early testing while still at their production facilities, effectively being organized as a distributed cyber-physical architecture. Such architecture demands robust, scalable, and automated testing processes to address geographic dispersion, heterogeneity, and strict real-time constraints~\cite{Broy2007,Chakraborty2016}.
As vehicle functionality increasingly relies on interconnected ECUs and over-the-air software updates, testing has become a significant cost driver in automotive development~\cite{Lami2016,Kushwaha2008}.

\begin{figure}[htbp]
    \centering
    \includegraphics[width=1\linewidth]{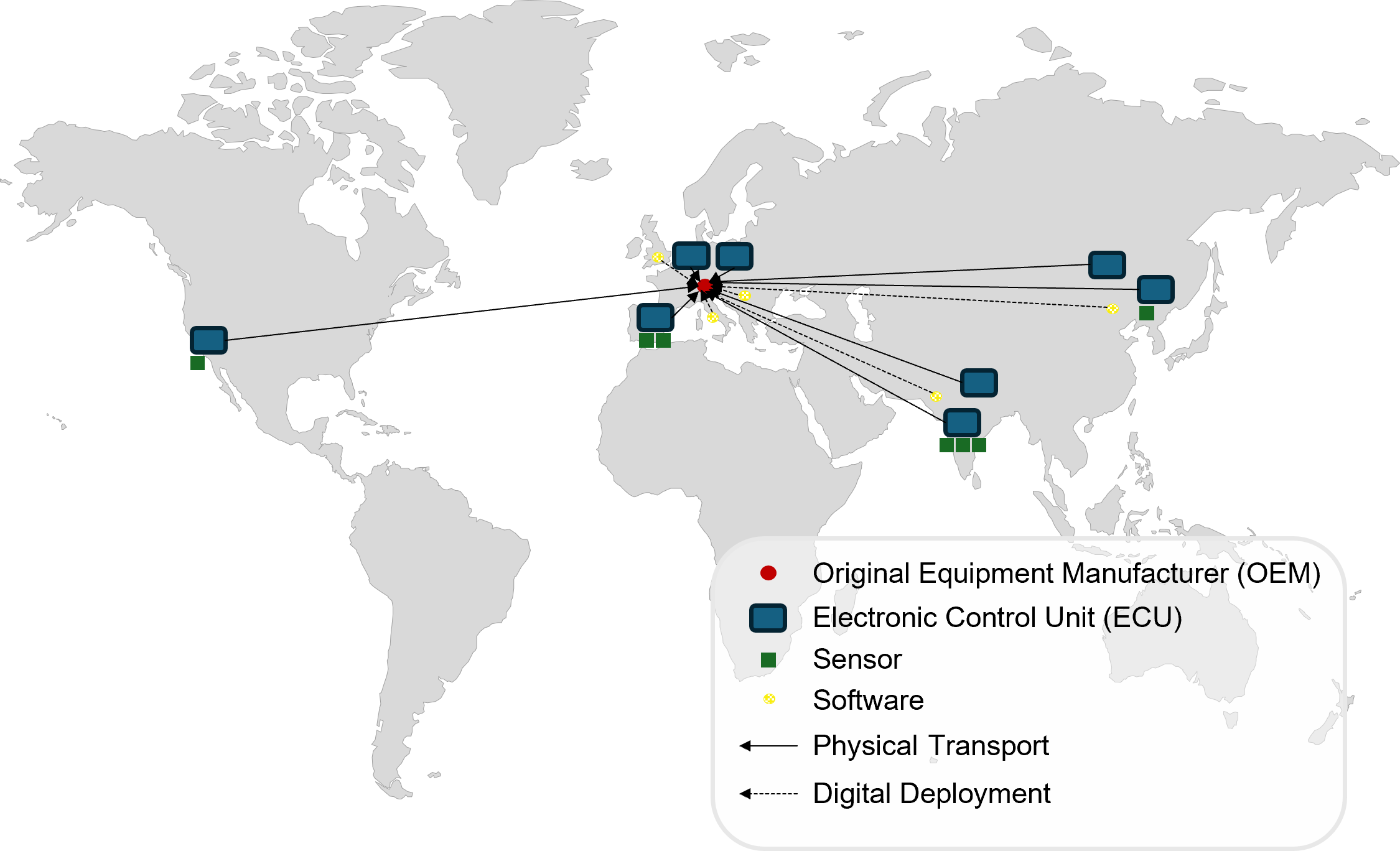}
    \caption{Global distribution of automotive testing components. Original Equipment Manufacturer (OEM) collaborates with geographically distributed suppliers providing ECUs, sensors, and software components.}
    \label{fig:worldmap}
\end{figure}

Ensuring correctness, safety, and performance across these decentralized systems is particularly challenging, especially given the organizational fragmentation in the automotive supply chain. Hardware and software components are often developed independently by globally distributed suppliers and integrated late at the Original Equipment Manufacturer’s (OEM) site (see Figure~\ref{fig:worldmap}). 

This asynchronous development process limits traceability and hinders test coordination. Validation is typically conducted by suppliers or external test labs, often with manually tracked test plans and inconsistent documentation. As a result, correlating test results across systems is difficult, valuable data remains underutilized, and comprehensive end-to-end validation is often postponed until all physical components are delivered. This delay increases iteration costs and prevents early-stage integration testing~\cite{Grimm2003,Pretschner2007,Rajkumar2010,Zyberaj2024}.

Traditional industry standards, such as ISO 26262~\cite{ISO26262-1-2018}, AUTOSAR~\cite{autosar}, and ASPICE~\cite{aspice}, provide useful guidance but do not fully support automated, infrastructure-aware testing for IoT-based architectures. Manual and semi-automat\-ed workflows still dominate, slowing feedback and integration into CI/CD pipelines.

In this paper, we introduce a fully automated testing pipeline for IoT-based automotive applications that generates Gherkin~\cite{gherkin} scenarios and executable test scripts from natural-language requirements and UML diagrams and links each scenario to its source requirement, using a large language model (LLM) and vision-language model (VLM) assistance with human-in-the-loop curation. The pipeline performs deployment-aware orchestration based on lightweight network profiling. It provides distributed execution via Eclipse openDuT~\cite{opendut}, emitting machine-readable results and Universally Unique Identifier (UUID)-based identifiers for end-to-end traceability and reproducibility. We implement a distributed Child Presence Detection System (CPDS) and demonstrate that it reduces manual authoring and deployment effort while enabling higher coverage across multiple nodes.

The remainder of the paper is structured as follows: Section~\ref{stateoftheart} provides background knowledge on testing of software and ECUs in the automotive industry. Section~\ref{approach} illustrates our proposed approach and implementation. Section~\ref{discussion} discusses results and evaluation. Section~\ref{relatedwork} reviews related works. Finally, Section~\ref{conclusion} summarizes our findings and outlines future research directions.

\section{Software and ECU Testing in the Automotive Industry}
\label{stateoftheart}

Continuous integration (CI) enables developers to integrate changes frequently and validate them through automated builds and tests, reducing integration risk and surfacing defects early~\cite{Duvall2013}. Testing is central to CI/CD pipelines and provides continuous feedback on software quality and readiness for deployment.
 
Automated tests at unit, component (integration), and acceptance levels are triggered by application, configuration, or environment changes. Unit tests give rapid feedback on isolated code. Component tests exercise interactions with databases, file systems, or external services. Acceptance tests validate the fulfillment of requirements and are often written in a behavior-driven development style using human-readable, executable specifications, for example, with Cucumber~\cite{cucumber} or JBehave~\cite{jbehave}. This alignment helps keep requirements, tests, and implementation synchronized and supports automatic regeneration of documentation with each build. Manual and exploratory testing remain important for usability checks, visual consistency, and open-ended scenarios that are hard to automate.

\begin{figure*}[htbp]
  \centering
  \includegraphics[width=\linewidth]{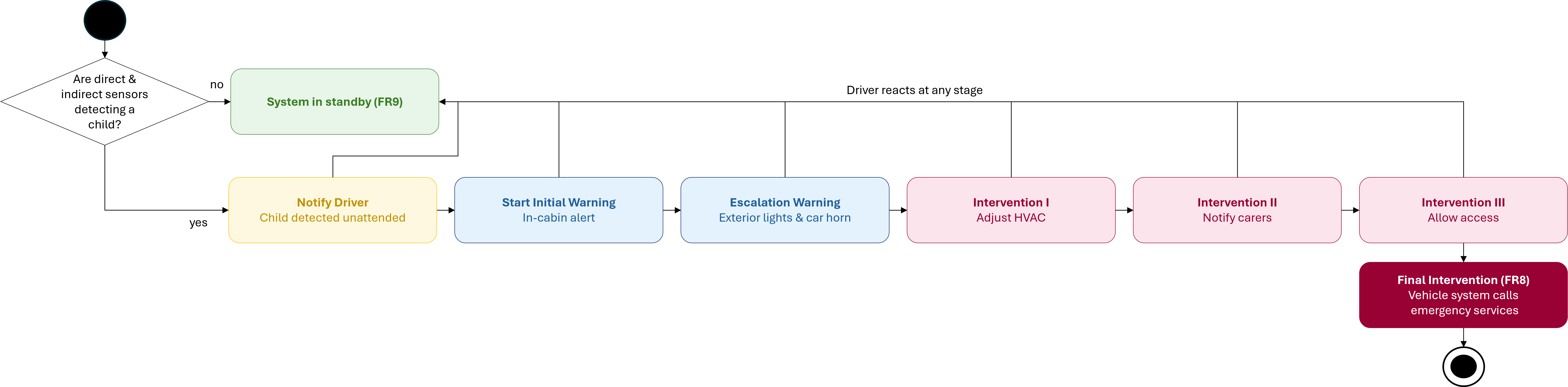}
  \caption{Overview of the simplified Child Presence Detection System (CPDS). Adapted from~\cite{zyberaj2025}.}
  \label{fig:child} 
\end{figure*}

Continuous delivery (CD) automates bringing software changes to a production-ready state, while continuous deployment also automates the release step. A deployment pipeline moves each change through build, test, and deploy stages and coordinates developers, testers, and operations~\cite{Humble2011}. A well-designed pipeline validates every change, improves release predictability, and increases transparency across the process. Standard CI/CD services (for example, Jenkins or GitLab CI) facilitate automated execution and management of these pipelines.

Adopting CI/CD and automated analysis in automotive contexts correlates with reduced defect rates, improved quality, and faster development cycles~\cite{Dakic2022,chatterjee2024}. Daki\'c et al. ~\cite{Dakic2022} report lower project costs and higher return on investment when these practices are integrated into automotive workflows. Beyond economic outcomes, research explores mechanisms for maintaining high coverage and intelligent test selection within CI/CD. For example, Fehlmann and Kranich~\cite{Fehlmann2020} propose a framework based on combinatory algebra and recursive arrow terms to recombine tests and map them to user stories via a coverage matrix, enabling relevance filtering and metric-driven suite management such as convergence gap and SVM-based optimization, which is relevant for Advanced Driver Assistance Systems (ADAS) systems with large operational spaces.

\section{Distributed Testing Pipeline}
\label{approach}
The automated testing of automotive software systems is organized here as an executable workflow. We implemented the pipeline as a working prototype and evaluated it on the CPDS use case, which reflects timing-sensitive, multi-sensor, distributed testing in an automotive IoT setting.

\subsection{Design of the Pipeline} 
\begin{figure*}[htbp]
  \centering
  \includegraphics[width=\linewidth]{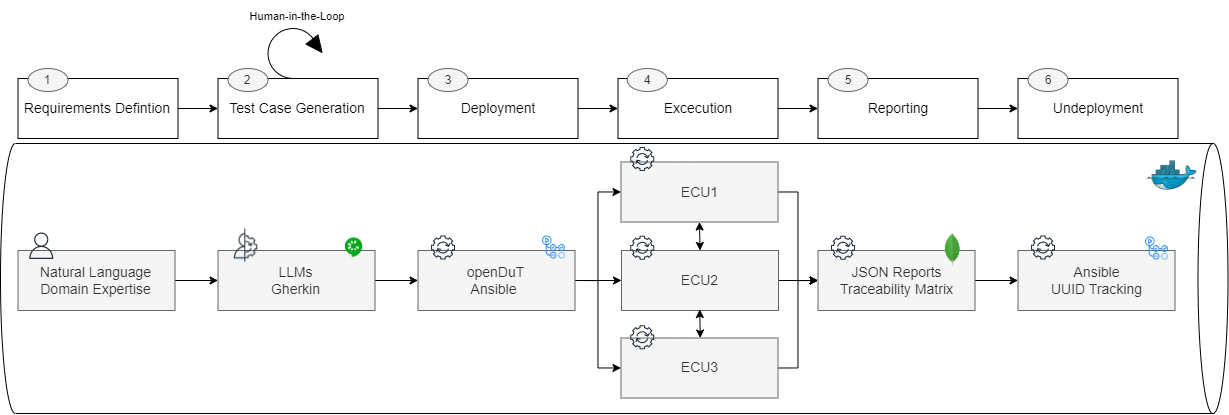}
  \caption{Overview of the distributed testing pipeline prototype, with main steps, supporting tools, and ECU communication via openDuT.}
  \label{fig:detailed_decentralized_pipeline} 
\end{figure*}

The pipeline consists of requirements definition, test case generation, deployment, execution, reporting, and undeployment. Figure~\ref{fig:detailed_decentralized_pipeline} shows the conceptual flow and the concrete tools used in the prototype, adapted from our prior work~\cite{zyberaj2025}. The goal is fully automated, distributed testing across ECUs with end-to-end traceability.

\textit{Step 1: Requirements definition.}
Requirements are authored in natural language and, where helpful, with diagrams such as UML activity or sequence diagrams that clarify interactions and timing. Clear requirements are critical in supplier settings where subsystems are developed independently. They form the basis for automated test case generation and later traceability.

\textit{Step 2: Test case generation.}
Test cases are automatically derived from natural-language requirements with LLM assistance and from UML diagrams with VLM assistance, with human-in-the-loop curation. We generate Gherkin scenarios and link each to its source requirement. A comprehensive evaluation of LLM and VLM models on this task and beyond (Vehicle Signal Specification signal mapping and Python test script generation) is provided in our companion study~\cite{zyberaj2026}, which benchmarks both commercial models (GPT-4o-mini, GPT-4.1, GPT-5, Gemini~2.5~Pro) and locally deployable alternatives (Qwen~2.5~VL~72B, Vicuna-7B) on the same CPDS case. The results show that a GPT-4 class model with Chain-of-Thought and Few-Shot prompting provides the best trade-off between generation quality and false-positive rate; accordingly, the present pipeline uses a GPT-4 class model for Gherkin generation. Details of the prompting strategy, guardrails, and iteration process are described in~\cite{zyberaj2026}. The resulting \texttt{.feature} files can embed deployment metadata, timing constraints, and scenario logic that guide script generation for distributed ECU testing.

\textit{Step 3: Deployment.}
Before deployment, the pipeline profiles device connectivity to account for heterogeneous networks. It gathers round-trip time, deviation, and packet loss and aggregates them into a normalized \textit{connection\_strength} score used to rank devices for placement and filtering. Deployment then uses scripts derived from the \texttt{.feature} files to configure sensors, start components, and set up logging. Orchestration is performed with Ansible~\cite{ansible} using an auto-generated inventory. Each deployment, execution, and undeployment task is tagged with a UUID to maintain traceability across CI/CD runs.

Distributed communication across ECUs is provided by Eclipse openDuT~\cite{opendut}, which offers secure connectivity and a hardware abstraction layer. In our setup, connectivity can be backed by systems such as NetBird~\cite{netbird} or WireGuard~\cite{wireguard}. This enables realistic OEM–supplier communication topologies and reduces integration overhead prior to in-vehicle tests.

\textit{Step 4: Execution.}
A central controller dispatches test instructions derived from the scenarios and coordinates simultaneous execution across ECUs. Tests are represented as JavaScript Object Notation (JSON) objects that define step logic, expected outcomes, and time constraints. The controller exposes a simple Representational State Transfer (REST) interface for triggering and monitoring runs, while lightweight gRPC messaging supports inter-ECU events for timely sensor–actuator signaling. Progress is streamed during execution and consolidated afterward into machine-readable summaries that support downstream analysis.

\textit{Step 5: Reporting.}
Results are aggregated into structured JSON with per-step outcomes, traces, and timings. Each item is mapped back to its Gherkin step and originating requirement to enable coverage analysis and faster root-cause investigation. A traceability matrix summarizes which scenarios validate which functional requirements.

\textit{Step 6: Undeployment.}
Post-test cleanup mirrors deployment. Inverse Ansible playbooks remove software components and temporary artifacts and reset devices to a clean state so subsequent runs start from a known baseline.

Table~\ref{tab:tools} provides an overview of all tools and components used in the pipeline, their respective pipeline steps, and the specific roles they play in the workflow.
\begin{table}[htbp]
  \centering
  \caption{Tools and components used in the pipeline}
  \label{tab:tools}
  \begin{tabularx}{\linewidth}{>{\raggedright\arraybackslash}p{2.2cm} 
    >{\raggedright\arraybackslash}p{2.0cm} Y}
    \toprule
    \textbf{Tool} & \textbf{Pipeline Step} & \textbf{Role} \\
    \midrule
    GPT-4 class LLM~\cite{zyberaj2026} 
      & Test case generation 
      & Generates Gherkin scenarios and executable Python test scripts 
        from natural-language requirements \\
    VLM (e.g., Gemini~2.5~Pro, Qwen~2.5~VL)~\cite{zyberaj2026} 
      & Test case generation 
      & Extracts signals and behavioral logic from UML diagrams to 
        enrich scenario generation \\
    Gherkin~\cite{gherkin} 
      & Test case definition 
      & Human-readable BDD scenario format linking requirements to 
        executable test logic \\
    Eclipse openDuT~\cite{opendut} 
      & Deployment, Execution, Management
      & Provides secure connectivity and hardware abstraction across 
        distributed ECUs \\
    Ansible~\cite{ansible}
      & Deployment, Undeployment 
      & Orchestrates automated deployment and teardown of software 
        components via auto-generated inventories \\
    NetBird / WireGuard ~\cite{netbird,wireguard}
      & Deployment 
      & Establishes encrypted overlay networks between ECUs to simulate 
        OEM--supplier topologies \\
    gRPC~\cite{grpc}
      & Execution 
      & Supports low-latency inter-ECU messaging for timely 
        sensor--actuator signaling \\
    REST
      & Execution 
      & Exposes the central controller interface for triggering and 
        monitoring test runs \\
    Python (unittest) 
      & Execution 
      & Runtime for generated test scripts; maps directly from Gherkin 
        steps to test methods \\
    JSON 
      & Reporting 
      & Universal artifact format for test definitions, execution 
        results, and traceability data \\
    MongoDB 
      & Reporting 
      & Stores test results, connectivity profiles, and pipeline 
        metadata during prototype runs \\
    GitHub Actions~\cite{githubactions}
      & CI/CD orchestration 
      & Triggers and manages the automated pipeline on self-hosted 
        runners \\
    \bottomrule
  \end{tabularx}
\end{table}

\subsection{CPDS Use Case Implementation} 
% Child detection
The Child Presence Detection System (CPDS) use case demonstrates the pipeline on a safety-critical scenario aligned with Euro NCAP guidance and related patents~\cite{euroncap,patent1,patent2}. Direct sensors, such as cabin camera, Ultra-Wideband (UWB) radar, and microphone, detect presence, while indirect sensors, such as seat pressure and belt sensors, provide corroboration. Detected events trigger escalating interventions up to emergency notification, as illustrated in Figure~\ref{fig:child}.
The CPDS prototype was implemented as a functional end-to-end demonstrator covering the full pipeline from requirements ingestion to post-test undeployment. Three ECUs were instantiated as software processes running on virtual nodes, each hosting a dedicated Python service that simulates its hardware component (door control, Heating, Ventilation, and Air Conditioning (HVAC), and infotainment). Sensor inputs were emulated via software-injected signals rather than physical hardware. The implementation was sufficient to validate all nine functional requirements and to exercise all six pipeline stages. It represents an early-integration prototype targeting OEM--supplier testing workflows rather than a production-grade deployment.

%requirement definition
We defined functional and technical requirements in Tables~\ref{tab:cpds_Frequirements} and \ref{tab:cpds_Trequirements}. The CPDS escalation logic is visualized in Figure~\ref{fig:child}. The combination of multi-sensor inputs, tight timing, and distributed orchestration makes CPDS a representative stress test for the pipeline.

\begin{table}[htbp]
  \centering
  \caption{Functional requirements for the CPDS}
  \label{tab:cpds_Frequirements}
  \begin{tabularx}{\linewidth}{p{1.8cm} Y}
    \toprule
    \textbf{Req. ID} & \textbf{Requirement description} \\
    \midrule
    FR1 & The CPDS shall detect an unattended child's presence using direct and indirect sensors. \\
    FR2 & The CPDS shall notify the driver of the forgotten child in the vehicle. \\
    FR3 & The CPDS shall initiate an initial warning upon detection if the driver does not react. \\
    FR4 & The CPDS shall escalate warnings using exterior lights and the horn if the initial warning fails. \\
    FR5 & The CPDS shall activate the HVAC to regulate interior conditions if previous warnings fail. \\
    FR6 & The CPDS shall notify all registered carers if the driver does not react. \\
    FR7 & The CPDS shall unlock vehicle doors to allow emergency access if prior interventions are unsuccessful. \\
    FR8 & The CPDS shall autonomously contact emergency services as a final safety measure. \\
    FR9 & The CPDS shall remain in standby if no unattended child is detected. \\
    \bottomrule
  \end{tabularx}
\end{table}

\begin{table}[h]
  \centering
  \caption{Technical requirements for the CPDS}
  \label{tab:cpds_Trequirements}
  \begin{tabularx}{\linewidth}{p{1.8cm} Y}
    \toprule
    \textbf{Req. ID} & \textbf{Requirement description} \\
    \midrule
    TR1 & Use camera, UWB radar, and microphone for direct detection. \\
    TR2 & Use seat sensors and door or window sensors for indirect detection. \\
    TR3 & Support gRPC for inter-ECU communication. \\
    TR4 & Allow real-time monitoring and status updates during escalation. \\
    TR5 & Enable automated generation and execution of deployment scripts. \\
    \bottomrule
  \end{tabularx}
\end{table}
%Test case generation
We used a GPT-4 class model to generate Gherkin scenarios from the requirements and then produced executable Python test scripts. Listing~\ref{lst:cpds_gherkin} shows a scenario for emergency notification, and Listing~\ref{lst:cpds_test} illustrates an executable excerpt.

\Needspace{8\baselineskip}
\begin{lstlisting}[style=gherkin-style, caption={Exemplary Gherkin test case for CPDS.}, label={lst:cpds_gherkin}]
Scenario: Child detected, no driver reaction, emergency services alerted
  Given the vehicle ignition is switched off
  And direct and indirect sensors detect a child alone in the vehicle
  When the system proceeds through all escalation steps
  And there is no reaction from the driver
  Then the vehicle system calls emergency services
  And allows emergency access to the vehicle
\end{lstlisting}

\begin{lstlisting}[style=python-style, caption={Executable test script excerpt for CPDS escalation.}, label={lst:cpds_test}]
def test_emergency_services_alerted_when_no_driver_reaction(self):
    self.vehicle_system.initiate_initial_warning()
    self.assertFalse(self.vehicle_system.driver_reacted())
    self.vehicle_system.escalate_warning()
    self.assertFalse(self.vehicle_system.driver_reacted())
    emergency_called = self.vehicle_system.call_emergency_services()
    self.assertTrue(emergency_called)
\end{lstlisting}

%deployment
Three ECUs were distributed across virtual or physical nodes interconnected through openDuT, simulating OEM–supplier arrangements. The devices and roles are listed in Table~\ref{tab:simulated_devices}. Connectivity profiling produced a normalized \textit{connection\_strength} score to guide placement, as shown in Listing~\ref{lst:network_scan}.

\begin{table*}[htbp]
  \centering
  \caption{Devices including IP address and core functionality}
  \label{tab:simulated_devices}
  \begin{tabularx}{\linewidth}{>{\raggedright\arraybackslash}p{2.2cm} >{\centering\arraybackslash}p{2.8cm} >{\raggedright\arraybackslash}p{3.2cm} Y}
    \toprule
    \textbf{Device name} & \textbf{IPv4 address} & \textbf{Software component} & \textbf{Functionality} \\
    \midrule
    ecu\_1 & 192.168.0.101 & door\_ecu.py & Doors \\
    ecu\_2 & 192.168.0.102 & ac\_ecu.py & Air conditioning \\
    ecu\_3 & 192.168.0.103 & infotainment\_ecu.py & Notifications and infotainment \\
    \bottomrule
  \end{tabularx}
\end{table*}

\Needspace{25\baselineskip}
%\begin{minipage}{\linewidth}
\begin{lstlisting}[style=json-style, caption={ECU connectivity profiling result.}, label={lst:network_scan}]
{
  "availability": 100,
  "availability_status": "available",
  "average_rtt": 0.06,
  "connection_strength": 0.999,
  "device_type": "heating_ventilation_and_airconditioning_ecu",
  "device_type_short": "HVAC",
  "hardware_version": "v1.0",
  "ip": "192.168.1.102",
  "max_rtt": 0.084,
  "mdev": 0.023,
  "min_rtt": 0.037,
  "name": "ecu_2",
  "normalized_packet_loss": 0.0,
  "sensors": [
    {"status": "available", "type": "temperature"},
    {"status": "available", "type": "status"}
  ],
  "software_packages": [],
  "status": "online"
}
\end{lstlisting}
%\end{minipage}

%execution
The CPDS scenarios are expressed as JSON to drive execution and to enable machine-readable reporting. A single representation is used throughout, ensuring consistent log parsing and traceability.

A central controller executes scenarios across ECUs, streams progress, and consolidates results. Execution results are stored as JSON with per-signal checks and per-step outcomes.
\Needspace{8\baselineskip}
\begin{lstlisting}[style=json-style, caption={Test JSON file for CPDS.}, label={lst:test_json}]
{
  "tests": [{
    "id": 1,
    "name": "Test car is locked",
    "teststeps": [1, 2, 3, 4],
    "logic_blocks": [{
      "id": 1,
      "logic_type": "TIME_CONSTRAINT",
      "condition_test_id": 1,
      "time_constraint_s": 3,
      "next_test_true": null,
      "next_test_false": null,
      "description": "Time constraint: 3 seconds passed"
    }]
  }]
}
\end{lstlisting}

%reporting
Table~\ref{tab:matrix} summarizes requirement coverage through a traceability matrix that maps scenarios to functional requirements. This supports faster root-cause analysis and requirements-driven regression selection.

\begin{lstlisting}[style=json-style, caption={Successful test result JSON for scenario ``Test car is locked''.}, label={lst:test_results}]
{
  "test_id": "1",
  "status": "completed",
  "result": {
    "status": "passed",
    "teststeps": [
      {"door_fl": {"expected_value": "0", "observed_value": "0", "status": "passed"}},
      {"door_fr": {"expected_value": "0", "observed_value": "0", "status": "passed"}},
      {"door_bl": {"expected_value": "0", "observed_value": "0", "status": "passed"}},
      {"door_br": {"expected_value": "0", "observed_value": "0", "status": "passed"}}
    ]
  }
}
\end{lstlisting}

\begin{table*}[htbp]
  \centering
  \caption{Traceability Matrix: Mapping Test Scenarios to Functional 
  Requirements}
  \label{tab:matrix}
  \renewcommand{\arraystretch}{1.2}
  \begin{tabularx}{\linewidth}{>{\raggedright\arraybackslash}p{5cm} 
    *{9}{>{\centering\arraybackslash}X}}
    \hline
    \textbf{Test Scenario} & \textbf{FR1} & \textbf{FR2} & \textbf{FR3} 
    & \textbf{FR4} & \textbf{FR5} & \textbf{FR6} & \textbf{FR7} 
    & \textbf{FR8} & \textbf{FR9} \\
    \hline
    Verification of unattended child detection 
      & \checkmark & & & & & & & & \\
    Child detected, driver reacts to initial warning 
      & \checkmark & \checkmark & & & & & & & \\
    Child detected, driver reacts after escalation warning 
      & \checkmark & \checkmark & \checkmark & & & & & & \\
    Child detected, driver reacts after climate control intervention 
      & \checkmark & \checkmark & \checkmark & \checkmark & & & & & \\
    Child detected, driver reacts after notifying carers 
      & \checkmark & \checkmark & \checkmark & \checkmark & \checkmark 
      & & & & \\
    Child detected, no driver reaction, emergency services alerted 
      & \checkmark & \checkmark & \checkmark & \checkmark & \checkmark 
      & \checkmark & \checkmark &\checkmark & \\
    No child detected 
      & \checkmark & & & & & & & & \checkmark \\
    \hline
  \end{tabularx}
\end{table*}

%undeployment
Post-test resource management uses inverse Ansible playbooks to remove components, reset configurations, and restore ECUs to a clean baseline so subsequent runs remain reproducible.

\section{Results and Discussion}
\label{discussion}
We evaluated the pipeline on the CPDS use case with three ECUs distributed across separate nodes. The evaluation covers three dimensions: test case generation quality, requirement coverage, and deployment feasibility.

\subsection{Test Case Generation Quality}
In the present pipeline, the 9~functional and 5~technical CPDS requirements (Tables~\ref{tab:cpds_Frequirements}  and~\ref{tab:cpds_Trequirements}) were deliberately scoped and phrased to avoid ambiguity, e.g., by avoiding compound \texttt{OR} conditions and ensuring each requirement maps to a single, observable behavior. Within this controlled set, the GPT-4 class model produced executable Gherkin scenarios for all 14 requirements without any manual correction, achieving 100\% generation accuracy. To contextualize this result against a more realistic, industry-scale requirement set, we draw on the companion evaluation in~\cite{zyberaj2026}, which applies the same generation approach to the full CPDS escalation logic expressed as 36 fine-grained natural-language requirements.

The CPDS requirement set in~\cite{zyberaj2026} expands the escalation logic into 36 fine-grained requirements; the generation quality figures cited here therefore reflect the requirement granularity used in that study and are directionally applicable to the coarser functional requirements used in the present pipeline. Of 36 natural-language CPDS requirements, 32 (89\%) were automatically transformed into executable Gherkin scenarios without modification. The remaining four required targeted corrections, primarily for requirements containing underspecified \texttt{OR} conditions or ambiguous edge cases~\cite{zyberaj2026}. Human review time averaged 1--5 minutes per requirement for Gherkin scenarios and 5--20 minutes per scenario for code adaptation, depending on complexity. These figures assume a reviewer with basic familiarity with the CPDS domain and practical test automation skills in Python~\cite{zyberaj2026}.

\subsection{Requirement Coverage}  
The traceability matrix in Table~\ref{tab:matrix} maps all seven test scenarios to their corresponding functional requirements. All nine functional requirements (FR1--FR9) are covered by at least one scenario, yielding full functional requirement coverage in the automated pipeline. Each requirement is traceable to its originating scenario via UUID-tagged JSON artifacts, supporting root-cause analysis and requirements-driven regression selection. Structured JSON-based test artifacts explicitly linked to requirements improve traceability and documentation consistency, overcoming common limitations of manual documentation.

\subsection{Deployment Feasibility}  
The successful implementation of the CPDS use case demonstrated the pipeline's practical applicability. 
Connectivity profiling produced a normalized connectivity score for each ECU prior to deployment; for instance, ecu\_2 achieved a \textit{connection\_strength} of 0.999, an average round-trip time of 0.06 ms, and zero packet loss (Listing~\ref{lst:network_scan}), enabling automated ranking and filtering of candidate nodes without manual configuration. Automated deployment via Ansible, test execution across all three ECUs, result aggregation, and post-test undeployment were all completed successfully, validating the core functionality and confirming the operational feasibility of the approach. By using Eclipse openDuT, the pipeline scales effectively across geographically distributed, heterogeneous ECU environments, simplifying the complexity of decentralized systems. Leveraging industry standards such as Ansible, REST, and gRPC enables the pipeline to integrate smoothly into existing automotive testing infrastructures, minimizing retooling effort.

\subsection{Limitations}  
Despite these results, several limitations arose during the implementation. The CI/CD pipeline implemented via GitHub Actions tightly couples the automation framework to GitHub, making it difficult to migrate to other version control systems or to pipeline tools such as Jenkins or GitLab CI. The self-hosted automation runners were developed explicitly for Windows environments, posing compatibility challenges with Linux systems. Additionally, the current implementation executes deployment, testing, and undeployment cycles sequentially, resulting in inefficient resource utilization and unnecessary wait times for developers. Scalability concerns were also noted regarding database performance and network constraints when expanding to larger ECU networks. MongoDB, used for prototyping, may encounter performance bottlenecks due to frequent concurrent access in large-scale deployments. Future improvements should focus on parallel execution capabilities, smarter scheduling algorithms informed by \textit{connection\_strength} and data dependencies, efficient resource management, enhanced network fault tolerance, and migration to time-series databases better suited for high-frequency IoT telemetry. The pipeline is currently aligned with functional safety verification as defined under ISO~26262. Extending the approach to cover the Safety of the Intended Functionality (SOTIF, ISO~21448)~\cite{iso21448}, for instance, by generating test scenarios for operational design domain boundary cases, remains a direction for future work. The inclusion of fault-injection scenarios, by simulating sensor failures or network dropout during execution, would further strengthen confidence in the pipeline's test validity and is planned as a future extension. To improve platform independence, replacing the current Windows-based self-hosted runners with a containerized execution environment (e.g., Docker-based runners) is a near-term engineering goal.

\section{Related Work} 
\label{relatedwork}
Several studies address CI implementations tailored to automotive software development. For instance, Nithnin et al.\ introduce the Contest Tool, an internal application designed to automate the downloading of source files, loading of software onto ECUs, execution of smoke tests, and generation of reports~\cite{Nithnin2023}. The tool enables rapid defect detection in ECU software, though it is not publicly available. A more comprehensive and unified CI infrastructure is described by Du et al.~\cite{Du2022}. Their proposal incorporates software emulation, fault injection, and Hardware-in-the-Loop (HiL) platforms. By integrating these stages into the CI pipeline, their approach enhances early fault detection and improves the overall reliability of system verification.

Obergfell et al.~\cite{Obergfell2019} suggest a transformation for automotive software engineering. They propose a dual-pipeline model that integrates architectural-level functionality with source-code-level implementation. Emphasizing agile methods, architectural extensibility, and enhanced supplier collaboration, the authors challenge the rigidity of the traditional V-model development approach. While the paper remains theoretical, it outlines key drivers and process-level shifts critical for OEMs adopting software-centric, continuous development practices. 

Broadening beyond purely automotive contexts but maintaining relevance to CI/CD practice, Soni~\cite{Soni2015} presents an end-to-end DevOps automation solution leveraging cloud technologies. The approach incorporates automated application lifecycle management, CI processes using Jenkins, configuration management via Chef, and infrastructure provisioning with VMware or public cloud environments\footnote{Chef: \url{https://www.chef.io/} and VMWare: \url{https://www.vmware.com/}.}, providing a valuable perspective applicable to automotive contexts seeking scalable, cloud-based solutions. 

Hilpert and Tetz~\cite{Hilpert2022} address the scalability challenges of automotive software testing by developing a fully automated prototype using Kubernetes-based cloud infrastructure. Their framework enables the scalable execution of Model-in-the-Loop (MiL) and Software-in-the-Loop (SiL) tests, demonstrating how cloud resources can be efficiently allocated for large-scale test automation. In the domain of model-based testing, Krej\v{c}\'{i} and Nov{\'a}k~\cite{Krejci2017} introduce a framework for automated test prioritization in distributed automotive systems. Their approach uses classification techniques to assign relevance scores to test cases, reducing redundant tests and focusing effort on the most critical tests for integration quality. Complementing this, Kim et al.~\cite{Kim2021} present the MAESTRO framework, which automates test generation from model specifications to achieve high coverage while minimizing manual input. Recent work has also addressed integrating machine learning into automotive CI/CD pipelines. Ferreira et al.~\cite{Ferreira2024} describe a Machine Learning Operations (MLOps) process for ML-enhanced automotive applications that automates the full machine learning lifecycle within a CI/CD context. Berglund et al.~\cite{Berglund2023} explore test maintenance strategies for machine learning systems in the automotive industry, identifying key challenges in test stability and reproducibility as models evolve. 

Compared to the approaches above, the present pipeline is distinguished by the combination of three properties that none of the reviewed works provide jointly: LLM and VLM assistance for test artifact generation starting directly from natural-language requirements and UML diagrams, deployment-aware orchestration driven by live network profiling, and full requirements-to-result traceability across a distributed, IoT-grade automotive ECU topology. Nithnin et al.\ and Du et al.\ support distributed ECU execution but rely on manually authored test cases without requirements-driven generation. Hilpert and Tetz achieve scalable multi-ECU execution in the cloud but do not address OEM--supplier geographic distribution, LLM-assisted generation, or end-to-end traceability. MAESTRO automates test generation from requirements, but requires pre-existing formal model specifications as input rather than operating from heterogeneous natural-language and diagram artifacts. Obergfell et al.\ address requirements-driven pipeline design at an architectural level, but remain theoretical and do not provide a working prototype.

\section{Conclusion and Outlook}
\label{conclusion}
In this paper, we present a fully automated, deployment-aware testing pipeline for IoT-based automotive applications and validated it on a distributed CPDS case. The pipeline converts natural-language requirements into Gherkin scenarios and executable scripts with LLM and VLM assistance and human-in-the-loop curation, orchestrates distributed execution via Eclipse openDuT, and emits machine-readable results with stable identifiers for end-to-end traceability. In our evaluation, it reduced manual authoring and deployment effort and supported higher coverage across multiple nodes. Within the controlled CPDS requirement set, the pipeline achieved 100\% Gherkin generation accuracy; at industry-scale requirement granularity, 89\% of scenarios required no manual correction~\cite{zyberaj2026}.
Distributed execution across three ECUs confirmed the pipeline's operational feasibility in an OEM-supplier network topology, with automated deployment and undeployment completing successfully across all test runs.

The current prototype has limits. It is realized on GitHub Actions with self-hosted Windows runners, executes scenarios sequentially, and uses a convenience data store that may not scale.

Future work focuses on scaling to larger multi-ECU settings, adding parallel execution with scheduling informed by \textit{connection\_strength} and data dependencies, and strengthening fault tolerance under variable network conditions. A systematic comparison of LLMs and VLMs for test generation is 
provided in our companion study~\cite{zyberaj2026}; future work will extend this to larger, multi-OEM requirement sets and diverse ECU configurations. To support reuse and collaboration, we will release the prototype and configuration templates as open source.
\bibliographystyle{splncs04}
 \bibliography{refs}

\end{document}